\documentclass[aps,twocolumn]{revtex4}
\usepackage[xdvi]{graphicx}
\usepackage{amsmath,revsymb}
\usepackage{dcolumn}

\begin{document}




\title{Theoretical interpretation of the experimental electronic
structure of lens shaped, self-assembled InAs/GaAs quantum dots}

\author{A.~J.~Williamson}
\author{L.W. Wang}
\author{Alex~Zunger}
\affiliation{National Renewable Energy Laboratory, Golden, CO 80401}

\begin{abstract}
We adopt an atomistic pseudopotential description of the electronic
structure of self-assembled, lens shaped InAs quantum dots within the
``linear combination of bulk bands'' method.  We present a detailed
comparison with experiment, including quantities such as the single
particle electron and hole energy level spacings, the excitonic band
gap, the electron-electron, hole-hole and electron hole Coulomb
energies and the optical polarization anisotropy.  We find a generally
good agreement, which is improved even further for a dot composition
where some Ga has diffused into the dots.
\end{abstract}
\maketitle


\section{Introduction: Using theory as a bridge between the
 structure and the electronic properties of quantum dots}
Self-assembled, Stranski-Krastanow (SK) grown semiconductor quantum
dots have recently received considerable attention as they exhibit a
rich spectrum of phenomena including
quantum-confinement\cite{hawrylak,gaponenko,mrs_feb98},
exchange-splittings\cite{landin98}, Coulomb
charging/blockade\cite{drexler94,medeiros95,fricke96,miller97,warburton97,schmidt97,warburton98,schmidt98,brunkov98}
and multi-exciton transitions\cite{landin98,dekel98}.  Over the past
few years a considerable number of high quality measurements of the
electronic level stucture of these dot systems have been performed,
using photoluminescence
(PL)\cite{schmidt97,schmidt98,yang99,yang2000,berryman97,itskevich98,itskevich99,fry2000},photoluminescence
luminescence excitation (PLE)\cite{landin98,dekel98},
capacitance\cite{drexler94,medeiros95,fricke96,brunkov98} and far
infra red (FIR)
spectroscopy\cite{fricke96,pan98,sauvage99,sauvage97,pan98:2}.  These
measurements have been able to determine the electronic level
structure to relatively high precision.  In parallel with these
measurements, several groups have also attempted to measure the
geometry and composition of these
dots\cite{yang99,yang2000,metzger99,garcia97,rubin96}.  So far,
however, these measurements have failed to provide details of the
shape, size, inhomogeneous strain and alloying profiles to a similar
level of accuracy to that in which the electronic structure has been
determined.  As a result, the size of the dots were often used as
adjustable parameters in models that fit experimental spectra.  For
example, using a single-band effective mass model, Dekel {\em
et. al}\cite{dekel98} defined an ``effective shape'' (cuboid) and
``effective dimension'' that reproduced the measured excitonic
transitions.  Similar ``parabolic dot'' models have been assumed by
Hawrylak {\em et. al}\cite{hawrylak}.

The accuracy of single-band and multi-band effective mass methods was
recently examined in a series of papers
\cite{wood96,fu_pressure,wang_cdse,pryor98:2,wang2000}.  In these
works, the shape, size and composition of nanostructures were
arbitrarily fixed, and the electronic structure was evaluated by
successively improving the basis set, starting from single-band
methods (effective-mass), going to six and eight band methods (k.p),
and finally, using a converged, multi-band approach (plane-wave
pseudopotentials).  It was found that conventional effective-mass and
k.p methods can sometimes significantly misrepresent the fully
converged results even when the shape, size and composition was given.
The observed discrepancies were both quantitative (such as band gap
values, level spacings, Coulomb energies) and qualitative (absence of
polarization anisotropy in square based pyramidal dots\cite{wang2000},
missing energy levels\cite{wang_cdse}).  As a result of these
limitations these methods may not offer a reliable bridge between the
electronic structure and atomic structure.

In this paper, we offer a bridge between recent measurements of the
{\em electronic structure} and measurements of the {\em atomic
structure} of the dots using accurate theoretical modeling.  Modeling
can determine if the calculated electronic structure resulting from an
assumed shape, size, strain and alloying profiles agrees with the
measured electronic structure or not.  A theory that can perform such
a ``bridging function'' must be accurate and reliable.  The
pseudopotential approach to this problem qualifies, in that any
discrepancy between the predicted and measured electronic properties
can be attributed to incorrectly assumed shape, size or alloying
profile.  We have studied a range of shapes, sizes and alloy profiles
and find that a lens-shaped InAs dot with an inhomogeneous Ga alloying
profile is in closest agreement with current measurements.  In the
following sections we attempt to provide a consistent theoretical
interpretation of numerous spectroscopic properties of InAs/GaAs dots.

\section{Outline of the Method of calculation}\label{methods}
We aim to calculate the energy associated with various electronic
excitations in InAs/GaAs quantum dots.  These energies can be
expressed as total energy differences and require four stages of
calculation:

(i) {\em Assume the shape, size and composition and compute the
equilibrium displacements:} We first construct a supercell containing
both the quantum dot and surrounding GaAs barrier material.  The
shape, size and composition profile are taken as input and
subsequently refined.  Sufficient GaAs barrier is used, so that when
periodic boundary conditions are applied to the system, the electronic
and strain interactions between dots in neighboring cells is
negligible.  The atomic positions within the supercell are then
relaxed by minimizing the strain energy described by an atomistic
force field\cite{keating,pryor98} including bond bending, bond
stretching and bond bending-bond stretching interactions (see
section~\ref{vff}).  An atomic force field is similar to continuum
elasticity approaches\cite{pryor98} in that both methods are based on
the elastic constants, $\{C_{ij}\}$, of the underlying bulk materials.
However, atomistic approaches are superior to continuum methods in two
ways, (a) they can contain anharmonic effects, and (b) they capture
the correct point group symmetry, e.g. the point group symmetry of a
square based, zinc blende pyramidal dot is $C_{2v}$, since the [110]
and [1$\overline{1}$0] directions are inequivalent while continuum
methods\cite{pryor98}, find $C_{4v}$.  More details of the atomistic
relaxation are given in section~\ref{vff}.

(ii) {\em Setup and solve the pseudopotential single-particle
equation:} A single-particle Schr\"odinger equation is set up at the
relaxed atomic positions, $\{{\bf R}_{n \alpha}\}$
\begin{equation}\label{ham}
\hat{H}\psi_i ({\bf r})=\{-{\beta\over 2} \nabla^2+ \sum_{n\alpha} \hat v_{\alpha} ({\bf r}
- {\bf R}_{n\alpha})\} \psi_i ({\bf r})= \epsilon_i\;\psi_i ({\bf r})
\;\;\; .
\end{equation}
The potential for the system is written as a sum of strain-dependent,
screened atomic pseudopotentials, $v_{\alpha}$, that are fit to bulk
properties extracted from experiment and first-principles calculations
(see section~\ref{sp}).  The Schr\"odinger equation is solved by
expanding $\psi$ in a linear combination of bulk states, $\phi_{nk}$,
from bands, $n$, and k-points, $k$,
\begin{equation}\label{lcbb_basis}
\psi_i({\bf r},\epsilon)=\sum_{n,k}c_{n,k}^{(i)}\;\phi_{nk}({\bf r},\epsilon) \;\;\;,
\end{equation}
taken at a few strain values.  The solution of
Eqs.~(\ref{ham}) and (\ref{lcbb_basis}) provides the level
structure and dipole transition matrix elements.  More details on the
solution of the Schr\"odinger equation are given in
section~\ref{lcbb}.

(iii) {\em Calculate the screened, inter-particle many-body
interactions:} The calculated single particle wavefunctions are used
to compute the electron-electron, electron-hole and hole-hole direct
, $J_{ee},J_{eh}, J_{hh}$, and exchange $K_{ee}, K_{eh}, K_{hh}$
Coulomb energies (see section~\ref{coulomb}).

(iv) {\em Calculate excitation energies as differences in total, many
particle energies:} For example, the difference between the total
energy $E_{11}[h_0^1e_0^1]$ of a dot with a hole in level $h_0$ and an
electron in level $e_0$ and the total energy $E_{00}[h_0^0e_0^0]$ of
the unexcited dot is
\begin{eqnarray}
E_{11}[h_0^1e_0^1]-E_{00}[h_0^0e_0^0] & = & \left(\epsilon_{e_0} -
\epsilon_{h_0}\right) \nonumber \\& - & J_{e_0h_0} +
2K_{e_0h_0}\delta_{S,0}\;\;\;,
\end{eqnarray}
where (in the absence of spin-orbit coupling) $\delta_{S,0}=1$ for
triplet states, and 0 for singlet states.  Analagous expressions exist
for electron-addition experiments (see section~\ref{coulomb}).

The main approximations involved in our method are: (a) the fit of the
pseudopotential to the experimental data of bulk materials is never
perfect (see Table~\ref{fit_table}) and (b) we neglect self-consistent
iterations in that we assume that the screened pseudopotential drawn
from a bulk calculation is appropriate for the dot.  Our numerical
convergence parameters are (i) the size of the GaAs barrier separating
periodic images of the dots, and (ii) the number of bulk wavefunctions
used in the LCBB expansion of the wavefunctions.  To examine the
effects of these approximations and convergences on the ultimate level
of accuracy that can be obtained with our methodology we have first
applied these methods to an InGaAs/GaAs quantum well(see
section~\ref{qw_test}), where experimental measurements of the shape,
size, composition and transition energies are more established (see
section~\ref{qw_test}).  We next describe the details of our method.
\begin{table}[hbt]
\caption{Fitted bulk electronic properties for GaAs and InAs using the
screened atomic pseudopotentials, in Eq.(\ref{v_loc}).  The
hydrostatic deformation potential of the band gap and $\Gamma_{15v}$
levels are denoted by $a_{gap}$ and $a_{\Gamma_{15v}}$.  The biaxial
deformation potential is denoted by $b$ and the spin-orbit splittings
at the $\Gamma_{15v}$ and $L_{1v}$ points are denoted by $\Delta_0$
and $\Delta_1$.}
\label{fit_table}
\begin{tabular*}{\linewidth}{@{\extracolsep{\fill}}ccccc}
\toprule
Property & \multicolumn{2}{c}{GaAs} & \multicolumn{2}{c}{InAs} \\
         &   EPM  & Expt\cite{bornstein} & EPM & Expt\cite{bornstein} \\
\colrule
$E_{gap}$ &      1.527	 &  1.52  &  0.424 &  0.42 \\   
$E_{X_{5v}}$ &    -2.697   & -2.96  & -2.330 & -2.40 \\
$E_{X_{1c}}$ &    1.981	 &  1.98  &  2.205 &  2.34 \\   
$E_{X_{3c}}$ &     2.52	 &  2.50  &  2.719 &  2.54 \\
$E_{L_{3v}}$ &    -1.01	 & -1.30  & -5.76  & -6.30 \\
$E_{L_{1c}}$ &     2.36	 &  1.81  &  1.668 &  1.71 \\   
$m_e^*$   &      0.066	 &  0.067 &   0.024 &  0.023 \\
$m^*_{hh}[100]$ & 0.342	 &  0.40  &   0.385 &  0.35 \\ 
$m^*_{hh}[111]$ & 0.866	 &  0.57  &   0.994 &  0.85 \\ 
$m^*_{lh}[100]$ & 0.093	 &  0.082 &   0.030 &  0.026 \\
$a_{gap}$        &    -7.88	 & -8.33  & -6.79  & -5.7 \\
$a_{\Gamma_{15v}}$ & -1.11 & -1.0 &  -0.826 & -1.0 \\
$b$        &    -1.559	 &  -1.7  &  -1.62 & -1.7 \\
$\Delta_0$ &      0.34 	 &  0.34  &    0.36 &  0.38 \\  
$\Delta_1$ &      0.177	 &  0.22  &    0.26 &  0.27 \\  
\botrule
\end{tabular*}
\end{table}

\section{Details of the Method of Calculations}
\subsection{Calculation of equilibrium atomic positions for a given shape}\label{vff}
To calculate the relaxed atomic positions within the supercell, we use
a generalization (G-VFF) of the original valence force field
(VFF)\cite{keating} model.  Our implementation of the VFF includes
bond stretching, bond angle bending and bond-length/bond-angle
interaction terms in the VFF Hamiltonian. This enables us to
accurately reproduce the $C_{11}$, $C_{12}$ and $C_{44}$ elastic
constants in a zincblende bulk material.  We have also included higher
order bond stretching terms, which lead to the correct dependence of
the Young's modulus with pressure.  The G-VFF total energy can be
expressed as:
\begin{eqnarray}\label{vff-eq}
E_{VFF} & = & \sum_i \sum_j^{nn_i} {3\over 8} [\alpha_{ij}^{(1)} \Delta
d_{ij}^2+ \alpha_{ij}^{(2)} \Delta d_{ij}^3 ] \nonumber \\
& + & \sum_i \sum_{k>j}^{nn_i} {3 \beta_{jik} \over 8 d_{ij}^0 d_{ik}^0 }
 [({\bf R}_j-{\bf R}_i)\cdot ({\bf R}_k-{\bf R}_i)\nonumber \\ & - & cos\theta^0_{jik} d_{ij}^0 d_{ik}^0]^2 \nonumber \\ 
& + & \sum_i \sum_{k>j}^{nn_i} {3 \sigma_{ijk}\over d_{ik}^0 } \Delta
d_{ij} [({\bf R}_j-{\bf R}_i)\cdot ({\bf R}_k-{\bf
R}_i)\nonumber \\ & - & cos\theta^0_{jik} d_{ij}^0 d_{ik}^0] \;\;\; ,
\end{eqnarray}
where $\Delta
d_{ij}^2=\left[[(R_i-R_j)^2-{d_{ij}^0}^2]/d_{ij}^0\right]^2$.  Here
${\bf R}_i$ is the coordinate of atom i and $d_{ij}^0$ is the ideal
(unrelaxed) bond distance between atom types of $i$ and $j$. Also,
$\theta^0_{jik}$ is the ideal (unrelaxed) angle of the bond angle
$j-i-k$.  The $\sum^{nn_i}$ denotes summation over the nearest
neighbors of atom $i$.  The bond stretching, bond angle bending, and
bond-length/bond-angle interaction coefficients $\alpha_{ij}^{(1)}
(\equiv \alpha)$, $\beta_{jik}$, $\sigma_{jik}$ are related to the
elastic constants in a pure zincblende structure in the following way,
\begin{eqnarray}\label{c11}
 C_{11}+2 C_{12} & = & {\sqrt{3}\over 4 d_0} (3 \alpha+\beta - 6
 \sigma) \nonumber \\
 C_{11}-C_{12} & = & {\sqrt{3}\over d_0} \beta  \nonumber \\
C_{44} & = & {\sqrt{3}\over d_0} {[(\alpha+\beta)
(\alpha\beta-\sigma^2)-2 \sigma^3 + 2 \alpha \beta \sigma ] \over
(\alpha+\beta+2 \sigma)^2 } \;\;\;.
\end{eqnarray}
The second-order bond stretching coefficient $\alpha^{(2)}$ is related
to the pressure derivative of the Young's modulus by ${d B\over d P}$,
where $B=(C_{11}+2 C_{12})/3$ is the Young's modulus.  Note that in
the standard\cite{keating} VFF which we have used
previously\cite{jkim98,williamson99,williamson98} the last terms of
Eq.(\ref{vff-eq}) are missing, so $\sigma=0$ in Eq.(\ref{c11}).  Thus
there were only {\em two} free parameters ($\alpha,\beta$) and
therefore three elastic constants could not, in general, be fit
exactly.  The G-VFF parameters and the resulting elastic constants are
shown in Table~\ref{vff-table} for GaAs and InAs crystals.  For an
InGaAs alloy system, the bond angle and bond-length/bond-angle
interaction parameters $\beta$, $\sigma$ for the mixed cation Ga-As-In
bond-angle are taken as the algebraic average of the In-As-In and
Ga-As-Ga values.  The ideal bond angle $\theta^0_{jik}$ is 109$^\circ$
for the pure zincblende crystal.  However, to satisfy Vegas's law for
the alloy volume, we find that it is necessary to use
$\theta^0_{Ga-As-In}=110.5^\circ$ for the cation mixed bond angle.
\begin{table}
\caption{Input G-VFF parameters $\alpha$, $\beta$, $\sigma$ to
Eq.(\ref{vff-eq}) and their resulting elastic constants $C_{11}$,
$C_{12}$ and $C_{44}$ ($10^3$ dyne/cm). }
\label{vff-table}
\begin{tabular*}{\linewidth}{@{\extracolsep{\fill}}lccccccc}
\toprule
& $\alpha$ & $\beta$ & $\sigma$  & $\alpha^{(2)}$ & $C_{11}$ & $C_{12}$ & $C_{44}$ \\ 
\colrule
GaAs & 32.153  &  9.370  & -4.099 & -105. &  12.11 &  5.48   &  6.04 \\
InAs & 21.674  &  5.760  & -5.753 & -112. &  8.33   & 4.53  &  3.80
\\
\botrule
\end{tabular*}
\end{table}

As a simple test of this G-VFF for alloy systems, we compared the
relaxed atomic positions from G-VFF with pseudopotential LDA results
for a (100) (GaAs)$_1$/(InAs)$_1$ superlattice where the $c/a$ ratio is
fixed to 1, but we allow energy minimizing changes in the overall
lattice constant ($a_{eq}$) and the atomic internal degrees of freedom
($u_{eq}$).  We find $a_{eq}^{LDA}=5.8612$ \AA~and
$u_{eq}^{LDA}=0.2305$, while the G-VFF results are
$a_{eq}^{G-VFF}=5.8611$ \AA~and $u_{eq}^{G-VFF}=0.2305$.  In
comparison the original VFF yields $a_{eq}^{VFF}=5.8476$ \AA~and
$u_{eq}^{VFF}=0.2303$.

\subsection{The Empirical Pseudopotential Hamiltonian}\label{sp}
We set up the single-particle Hamiltonian as
\begin{equation}\label{hamiltonian}
\hat{H}=-{\beta\over 2} \nabla^2+ \sum_{n\alpha} \hat v_{\alpha} ({\bf
r} - {\bf R}_{n\alpha}) \;\;\; ,
\end{equation}
where ${\bf R}_{n\alpha}$ is the G-VFF relaxed position of the
n$^{th}$ atom of type $\alpha$.  Here $\hat v_{\alpha}({\bf r})$ is a
screened empirical pseudopotential for atomic type $\alpha$.  It
contains a local part and a nonlocal, spin-orbit interaction part.

The local potential part is designed to include dependence on the
local hydrostatic strain Tr$(\epsilon)$:
\begin{equation}\label{v_loc}
v_{\alpha}^{loc}(r;\epsilon) = v_{\alpha}^{eq}(r;0) [1+\gamma_\alpha
 {\rm Tr} (\epsilon)] \;\;\;,
\end{equation}
where the $\gamma_\alpha$ is a fitting parameter.  The zero strain
potential $v_{\alpha}^{eq}(r;0)$ is expressed in reciprocal space q as
\begin{equation}\label{v_q}
v(q)=a_0 (q^2-a_1)/[a_2 e^{a_3 q^2}-1] \;\;\;.
\end{equation}
The local hydrostatic strain $Tr(\epsilon)$ for a given atom at ${\bf
R}$ is defined as $\Omega_R/\Omega_0-1$, where $\Omega_R$ is the
volume of the tetrahedron formed by the four atoms bonded to the atom
at ${\bf R}$.  $\Omega_0$ is the volume of that tetrahedron in the
unstrained condition.  The need for explicit dependence of the atomic
pseudopotential on strain in Eq.(\ref{v_loc}) results from the
following: While the description in Eq.(\ref{hamiltonian}) of the
total pseudopotential as a superposition of atomic potentials situated
at specific sites, $\{R_{n\alpha}\}$, does capture the correct local
symmetries in the system, the absence of a self-consistent treatment
of the Schr\"odinger equation deprives the potential from changing in
response to strain.  In the absence of a strain-dependent term, the
volume dependence of the energy of the bulk valence band maximum is
incorrect.  While self-consistent descriptions show that the volume
deformation potential $a_v=dE_v/d \ln \Omega$ of the valence band
maximum is {\em negative} for GaAs, GaSb, InAs, InSb and for all II-VI
this qualitative behavior can not be obtained by a non-self-consistent
calculation that lacks a strain dependent pseudopotential.

The nonlocal part of the potential describes the spin-orbit
interaction,
\begin{eqnarray}
H_{so} & = & \sum_{n\alpha} \hat V^{so}_{\alpha}(R_{n\alpha})
\\ & \equiv &
\sum_{n\alpha} \sum_l V^{so}_{l,\alpha}(r-R_{n\alpha})
|l\rangle_{R_{n\alpha}} {\bf L}\cdot {\bf S} \langle l
|_{R_{n\alpha}} \nonumber  \;\;\;,
\end{eqnarray}

where $|l\rangle_{R_{n\alpha}}$ is a projector of angular momentum $l$
centered at $R_{n\alpha}$, ${\bf L}$ is the spatial angular momentum
operator, ${\bf S}$ is the Dirac spin operator, and
$V^{so}_{l,\alpha}(r)$ is a potential describing the spin-orbit
interaction.  

In Eq(\ref{hamiltonian}), the kinetic energy of the electrons has been
scaled by a factor of $\beta$.  The origin of this term is as follows:
In an accurate description of the crystal band structure, such as the
GW method\cite{hedin99}, a general, spatially non-local potential,
$V(r,r')$, is needed to describe the self-energy term.  In the absence
of such a term the occupied band width of an inhomogeneous electron
gas is too large compared to the exact many-body result.  To a first
approximation, however, the leading effects of this non-local
potential, $V(r,r')$, can be represented by scaling the kinetic
energy.  This can be seen by Fourier transforming $V(r,r')$ in
reciprocal space, $q$, then making a Taylor expansion of $q$ about
zero.  We find that the introduction of such a kinetic energy scaling,
$\beta$ permits a simultaneous fit of both the effective masses and
energy gaps.  In this study, we fit $\beta=1.23$ for both GaAs and
InAs.

The pseudopotential parameters in Eqs(\ref{v_loc}) and (\ref{v_q})
were fitted to the bulk band structures, experimental deformation
potentials and effective masses and first-principles calculations of
the valence band offsets of of GaAs and InAs. The alloy bowing
parameter for the GaInAs band gap (0.6 eV) is also fitted.  The
pseudopotential parameters are given in Table~\ref{param_table} and
their fitted properties are given in
Table~\ref{fit_table}\cite{bornstein}.  We see that unlike the LDA,
here we accurately reproduce the bulk band gaps and the bulk effective
masses.  One significant difference in our parameter set, to that used
in conventional k.p studies, is our choice of a negative magnitude for
the valence band deformation potential, $a_v$, which we have obtained
from LAPW calculations\cite{franceschetti94}.
\begin{table}[hbt]
\caption{Parameters for the GaAs and InAs screened atomic
pseudopotentials, in Eq.(\ref{v_loc}).  This potential requires a
plane wave cutoff of 5 Ryd.}
\label{param_table}
\begin{tabular*}{\linewidth}{@{\extracolsep{\fill}}ccccc}
\toprule
Parameter & In & Ga & As (InAs) & As (GaAs) \\
\toprule
a$_0$ & 644.13 & 432960 & 26.468 & 10.933 \\
a$_1$ & 1.5126 & 1.7842 & 3.0313 & 3.0905 \\
a$_2$ & 15.201 & 18880  & 1.2464 & 1.1040 \\
a$_3$ & 0.35374 & 0.20810 & 0.42129 & 0.23304 \\
a$_4$ & 2.1821 & 2.5639 & 0.0 & 0.0 \\
\botrule
\end{tabular*}
\end{table}

The present InAs and GaAs pseudopotentials have been systematically
improved relative to our previous InAs and GaAs
potentials\cite{jkim98,williamson99,williamson98:2,wang99,wang99:2},
although the functional form has remained the same.  Firstly, the
pseudopotentials for InAs and GaAs used in
Ref.\cite{jkim98,wang99} did not include the spin-orbit
interaction.  In
Refs.\cite{williamson98:2,williamson99,wang99:2} we used
potentials that included the spin-orbit interaction, but were not able
to simultaneously, accurately fit the electron effective and the zone
center band gap, due to the lack of the above $\beta$ parameter.  The
potential used here is identical to that used in
Refs.\cite{wang99,wang2000}.

\subsection{Calculating the single particle eigenstates}\label{lcbb}
One could use a straight forward expansion of the single particle
wavefunctions in a plane wave basis set, as we have previously done in
Refs.\cite{williamson98,williamson99,jkim98}.  However, as was
shown in Refs.\cite{wang99,wang2000,wang97:2}, a more economical
representation is to use the Linear Combination of Bulk Bands (LCBB)
method\cite{wang99,wang2000,wang97:2}.  Within the LCBB the
eigenstates of the pseudopotential Hamiltonian are expanded in a basis
of bulk Bloch orbitals
\begin{equation}\label{basis}
\psi_i({\bf
r})=\sum_s\sum_{n,k}c_{s,n,k}^{(i)}\;u_{s,n,k}({\bf
r})\;e^{i{\bf k.r}} \;\;\;,
\end{equation}
where $u_{s,n,k}({\bf r})$ is the cell periodic part of the bulk Bloch
wavefunction for structure, $s$, at the $n^{th}$ band and the k$^{th}$
k-point, $k$.  These states form a physically more intuitive basis
than traditional plane waves therefore the number of bands and
k-points can be significantly reduced to keep only the physically
important bands and k-points (around the $\Gamma$ point in this case).
This method was recently generalized to strained semiconductor
heterostructure systems\cite{wang99} and to include to spin-orbit
interaction\cite{wang99:2}.  In this paper use an LCBB basis derived
from four structures, $s$.  These structures are (i) unstrained, bulk
InAs at zero pressure, (ii) unstrained, bulk GaAs at zero pressure,
(iii) bulk InAs subjected to the strain value in the center of the
InAs dot, and (iv) bulk InAs subjected to the strain value at the tip
of the InAs dot.  By interpolating the strain profile between these
four structures, the basis is able to accurately describe all the
strain in the system.  The wavevectors, $\{k\}$, used here include all
allowed values within $16\pi/L$ of the zone center, where $L$ is the
supercell size.  For calculations of electron states, the band
indices, $n$, include only the band around the $\Gamma_{1c}$ point.
For the hole states we also include the three bands around the
$\Gamma_{15v}$ point.  This basis set produces single particle
energies that are converged with respect to basis size, to within 1
meV.

\subsection{Constructing the energies of different electronic
configurations}\label{coulomb}
Using screened Hartree Fock theory, the energy associated with loading
$N$ electrons into a quantum dot can be
expressed\cite{franceschetti2000} as
\begin{equation}\label{total_energy}
E_N = \sum_i(\epsilon_i+\Sigma_i^{pol})n_i +
\sum_{i<j}(J^{ee}_{ij}-K^{ee}_{ij})n_in_j \;\;\;,
\end{equation}
where $\epsilon_i$ are the single-particle energies and
$\Sigma_i^{pol}$ are the polarization self-energies of the $i^{th}$
electron state , $J^{ee}_{ij}$ and $K^{ee}_{ij}$ are the direct and
exchange Coulomb integrals between the $i^{th}$ and $j^{th}$
electronic states and $n_i$ are the occupation numbers
($\sum_in_i=N$).  As shown in Ref.\cite{franceschetti2000}, for
free standing, colloidal quantum dots the dielectric constant inside
the dot is dramatically different to that outside (vacuum) and hence
the polarization self-energy, $\Sigma_i^{pol}$, is very significant
($\sim$1 eV).  For self assembled InAs dots embedded in GaAs, the
dielectric constants of InAs and GaAs are similar
($\epsilon_{\infty}=12.3,10.6$) and we calculate this term as $\sim$1
meV, and hence we choose to neglect it here.  The direct and exchange
Coulomb energies, are defined as
\begin{eqnarray}\label{Jeh}
J_{ij} & = & \int\int\frac{|\psi_i({\bf r}_1)|^2\;|\psi_j({\bf
r}_2)|^2}{\overline{\epsilon}({\bf r}_1-{\bf r}_2)|{\bf r}_1-{\bf
r}_2|} d{\bf r}_1 d{\bf r}_2 \nonumber \\ \nonumber \\ 
K_{ij} & = & \int\int\frac{\psi_i^*({\bf r}_1)\;\psi_i({\bf
r}_2)\;\psi_j^*({\bf r}_2)\;\psi_j({\bf r}_1)}{\overline{\epsilon}({\bf
r}_1-{\bf r}_2)|{\bf r}_1-{\bf r}_2|} d{\bf r}_1 d{\bf r}_2 \;\;\;,
\end{eqnarray}
where $\overline{\epsilon}$ is a phenomenological, screened dielectric
function\cite{williamson98:2} containing a Thomas Fermi electronic
component and an ionic component from Ref.\cite{haken}.  Our
exchange automatically includes both short and long range components.

Denoting electron levels as $e_0,e_1,e_2$..., hole levels as
$h_0,h_1,h_2$... and the number of electrons and holes as $N$ and $M$,
the total energy, $E_{MN}$, is 
\begin{eqnarray}\label{total_energy2}
E_{MN} & = & \sum_i-\epsilon_{h_i}m_i + \sum_{i<j}(J^{hh}_{ij}-K_{ij}^{hh})m_im_j
\nonumber \\
& + & \sum_i\epsilon_{e_i}n_i + \sum_{i<j}(J^{ee}_{ij}-K^{ee}_{ij})n_in_j \nonumber \\
& - & \sum_{ij}(J^{eh}_{ij}-K_{ij}^{eh})n_im_j \;\;\;,
\end{eqnarray}
where $n_i$ and $m_i$ are the electron and hole occupation numbers
respectively and $\sum_in_i=N$ and $\sum_im_i=M$.  Using
Eq.(\ref{total_energy2}), in the strong confinement regime where
kinetic energy effects dominate over the effects of exchange and
correlation, an exciton involving electrons excited from hole state
$i$ to electron state $j$ can be expressed
as
\begin{equation}\label{exciton}
E_{ij}^{exciton} = \left(\epsilon_{e_j} - \epsilon_{h_i}\right) -
J^{eh}_{ji} + K^{eh}_{ji}\delta_{S,0}\;\;\;.
\end{equation}
To study charged dots, if one assumes the electron states are occupied
in order of increasing energy (Aufbau principle), the total energy of
a dot charged with $N$ electrons, $E_{0N}$, is
\begin{eqnarray}\label{ee1}
E_{00}\;[e_0^0] & = & 0 \nonumber \\ E_{01}\;[e_0^1] &
= & \epsilon_{e_0} \nonumber \\ 
E_{02}\;[e_0^2] & = & 2\epsilon_{e_0} + J_{e_0,e_0} \nonumber \\
E_{03}\;[e_0^2e_1^1] & = & \left(2\epsilon_{e_0} + \epsilon_{e_1} \right)+
\left[ J_{e_0,e_0} + 2J_{e_0,e_1} \right]- K_{e_0,e_1} \nonumber \\
E_{04}\;[e_0^2e_1^2] & = & \left(2\epsilon_{e_0} + 2\epsilon_{e_1} \right)+
\left[ J_{e_0,e_0} + J_{e_1,e_1} + 4J_{e_0,e_1}\right] \nonumber \\
& - & 2K_{e_0,e_1} \;\;\;.
\end{eqnarray}
As indicated in section~\ref{methods}, our wavefunctions,
$\{\psi_i\}$, are not iterated to self-consistency.  This affects the
magnitude of the direct and exchange Coulomb integrals.  We have
previously examined the accuracy of this perturbative treatment for
colloidal InAs dots by comparing the non-self-consistent Coulomb
energy with that obtained self consistently\cite{franceschetti97}.
The differences were negligible.

\subsection{Quantum well tests}\label{qw_test}
\begin{figure}[h]
\includegraphics[width=\linewidth]{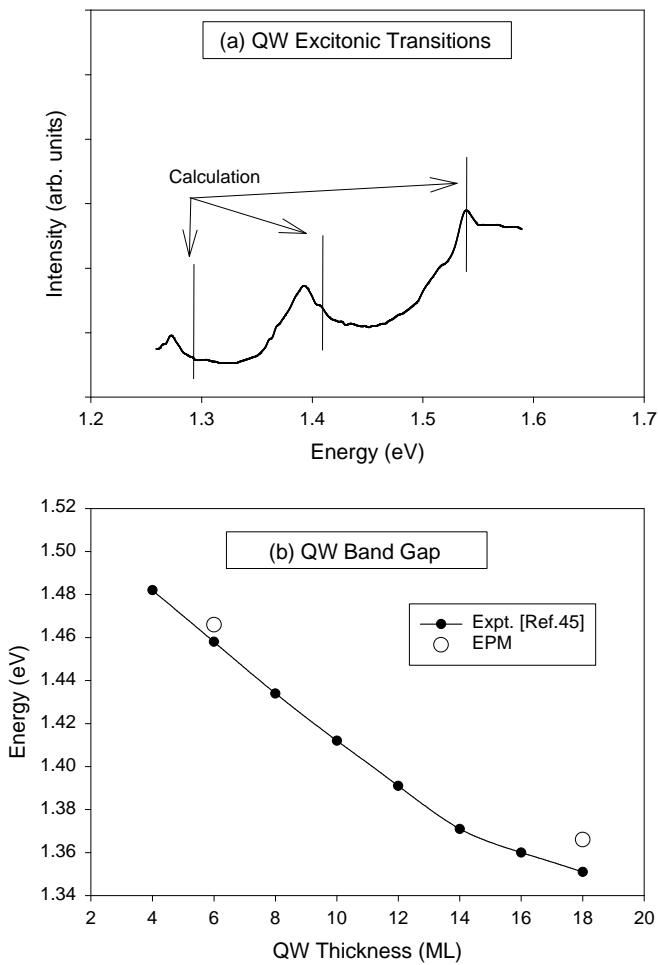}
\caption{(a) A comparison of EPM calculated and measured electron to
heavy hole transition energies in a 96 \AA~In$_{0.24}$Ga$_{0.76}$As
quantum well embedded inside a GaAs matrix.  The vertical lines mark
the positions of the EPM calculated transitions.  (b) The calculated
band gap of an In$_{0.22}$Ga$_{0.78}$As quantum well as a function of
its thickness.}
\label{qw_fig}
\end{figure}

To test the above methods, we first calculated the energy levels in a
quantum well, and compared the results with experiment. In
Fig.~\ref{qw_fig}(a), we compare the calculated electron-heavy hole
transition energies for a 96 \AA~In$_{0.24}$Ga$_{0.76}$As quantum
well inside a GaAs matrix.  The peaks in the experimental spectra
occur\cite{gershoni89} at 1.275, 1.395 and 1.538 eV.  Our calculated
transitions occur at 1.290, 1.404 and 1.545 eV
respectively. Figure~\ref{qw_fig}(b) compares the band gap of a
In$_{0.22}$Ga$_{0.78}$As quantum well as a function of its thickness.
The measured band gaps\cite{laymarie95} for quantum wells with
thicknesses of 6 and 18 ML are 1.458 and 1.351 eV.  Our calculated
values are 1.466 and 1.366 eV.

\section{Physical quantities to compare with experiment}\label{quantities}
\begin{figure}[h]
\includegraphics[width=\linewidth]{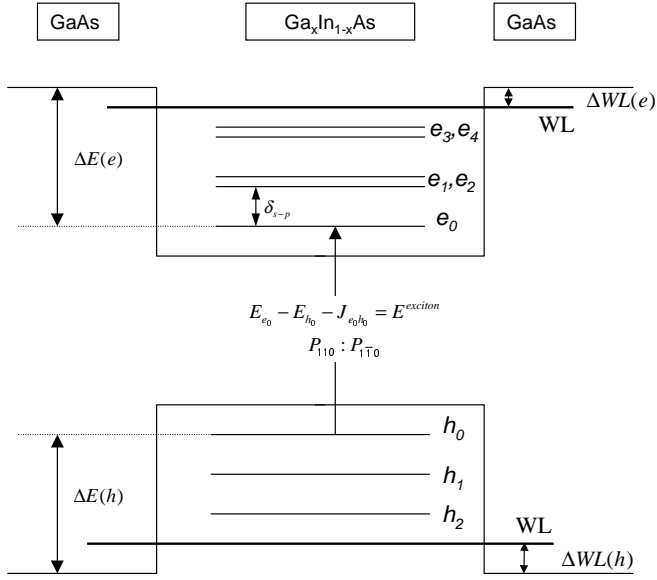}
\caption{Schematic single particle electron and hole energy levels for lens
shaped, InAs quantum dots embedded within GaAs.}
\label{level_fig}
\end{figure}

The quantities we use to characterize the electronic structure are
illustrated in Fig.~\ref{level_fig} which shows a schematic layout of
the electron and hole single-particle energy levels in a quantum dot.
Assuming that all levels are spatially nondegenerate (thus having only
spin degeneracy), we mark the electron levels as $e_0, e_1,
e_2$.... and the hole levels as $h_0, h_1, h_2$.  The level $e_0$ is
sometimes called ``s-like'', whereas $e_1$ and $e_2$ are called
``p-like'', and $e_3$ and $e_4$ are called ``d-like''.  Since the GaAs
environment of the InAs dots is largely unstrained, it is convenient
to set as a reference energy the VBM of GaAs as $E=0$, and the CBM of
GaAs as E=1520 meV.  All energy levels can be referenced with respect
to these band edges.

\noindent For the {\em electron levels}, the quantities that we
consider are:

(i) The number of dot-confined electron states, $N_e$.

(ii) The spacing $\delta_{sp}=\epsilon_{e_1}-\epsilon_{e_0}$ between ``s-like''
and ``p-like'' electron states.

(iii) The splitting $\delta_{pp}=\epsilon_{e_2}-\epsilon_{e_1}$ between the
``p-like'' electron states.

(iv) The spacing $\delta_{pd}=\epsilon_{e_3}-\epsilon_{e_2}$ between ``p-like''
and ``d-like'' electron states.

(v) The ``binding energy'' of the first electron level, $e_0$, with
respect to the GaAs conduction band minimum, $\Delta E(e)=E_{GaAs,
CBM}-\epsilon_{e_0}$.

(vi) The position of the bottom of the band for the 2D InAs ``wetting
layer'' (WL) with respect to the GaAs CBM, $\Delta E_{WL}^{(e)}=E_{GaAs,
CBM}-E^{(e)}_{WL}$.

(vii) Inter-electron direct $J_{e_i,e_j}^{ee}$ and exchange
$K_{e_i,e_j}^{ee}$ Coulomb energies.

\noindent For the {\em hole levels} we consider are:

(i) The number, $N_h$, of dot-confined hole states.

(ii) The intra-band spacings of the hole levels,
$\delta_{ij}^{(h)}=\epsilon_{h_j}-\epsilon_{h_i}$.

(iii) The ``binding energy'' of the first hole level, $h_0$, with
respect to the GaAs valence band maximum, $-\Delta E(h)=E_{GaAs,
VBM}+\epsilon_{h_0}$.

(iv) The position of the top of the band for the 2D InAs ``wetting
layer'' (WL) with respect to the GaAs VBM, $\Delta E_{WL}^{(h)}=-E_{GaAs,
VBM}+E^{(h)}_{WL}$.

(v) Inter-hole direct $J_{h_i,h_j}^{hh}$ and exchange
$K_{h_i,h_j}^{hh}$ Coulomb energies.

\noindent Finally, for the {\em recombination of electrons and holes},
we consider:

(i) The excitonic energies, $E_{ij}^{exciton}$, as defined in
Eq.(\ref{exciton}).  By subtracting calculated values for the single
particle energies $\epsilon_{e_j}$ and $\epsilon_{h_i}$ from measured
optical excitation energies one can estimate the electron-hole direct
Coulomb energies $J_{h_ie_j}$.

(ii) The ratio of absorption intensities for light polarized along
[110] and [1$\overline{1}$0] directions, defined as
\begin{equation}\label{lambda}
\lambda=\frac{P_{[110]}}{P_{[1\overline{1}0]}} =
\frac{<\psi_{e_0}|r_{[110]}|\psi_{h_0>^2}}{<\psi_{e_0}|r_{[1\overline{1}0]}|\psi_{h_0}>^2} \;\;\; .
\end{equation}
This ratio can deviate from unity due to three reasons; (a) The dots
has different dimensions in the [110] and [1$\overline{1}$0]
directions.  We refer to this as the the ``geometric factor''.  (b)
The atomistic zincblende symmetry makes the two directions symmetry
inequivalent even if the lengths along the two directions are equal.
We refer to this as the ``atomic symmetry factor''.  One manifestation
of this affect is that if the strain is calculated atomistically, it
is different in the two directions even in the absence of a geometric
factor\cite{pryor98}.  (c) A piezoelectric field that breaks the
symmetry.  Previous studies\cite{yang2000} have shown that this effect
is negligible in InAs/GaAs dots so we will neglect it here.  k.p
calculations neglect the ``atomic symmetry'' factor (except for the
small effect of strain asymmetry), but retain the ``geometric
factor''.  Pseudopotential calculations retain both effects.  For
example, in a {\em square} based pyramid (where by definition the
``geometric factor'' does not contribute), k.p produces $\lambda=1$,
while pseudopotential theory gives $\lambda=1.2$ (see
Table~\ref{results_table}).  This shows that there is not a simple
mapping from dot shape to polarization anisotropy, $\lambda$.

(iii) Excitonic dipole: As the center of the electron and hole wavefunctions
do not exactly coincide with each other, it
is possible that an exciton will exhibit a detectable dipole moment,
\begin{equation}\label{dipole}
d_{h_i,e_j}=\left< \psi_{h_i}|\hat{r}|\psi_{h_i}\right> - \left<
\psi_{e_j}|\hat{r}|\psi_{e_j}\right>  \;\;\;.
\end{equation}

The quantities defined above characterize the electronic structure.
Next, in section~\ref{results}, we will next provide all of these
quantities from our calculations, and then in section~\ref{expt} we
will extract measured values of these quantities from the available
experiments.

\section{Theoretical Results}\label{results}

\begin{figure}[h]
\includegraphics[width=\linewidth]{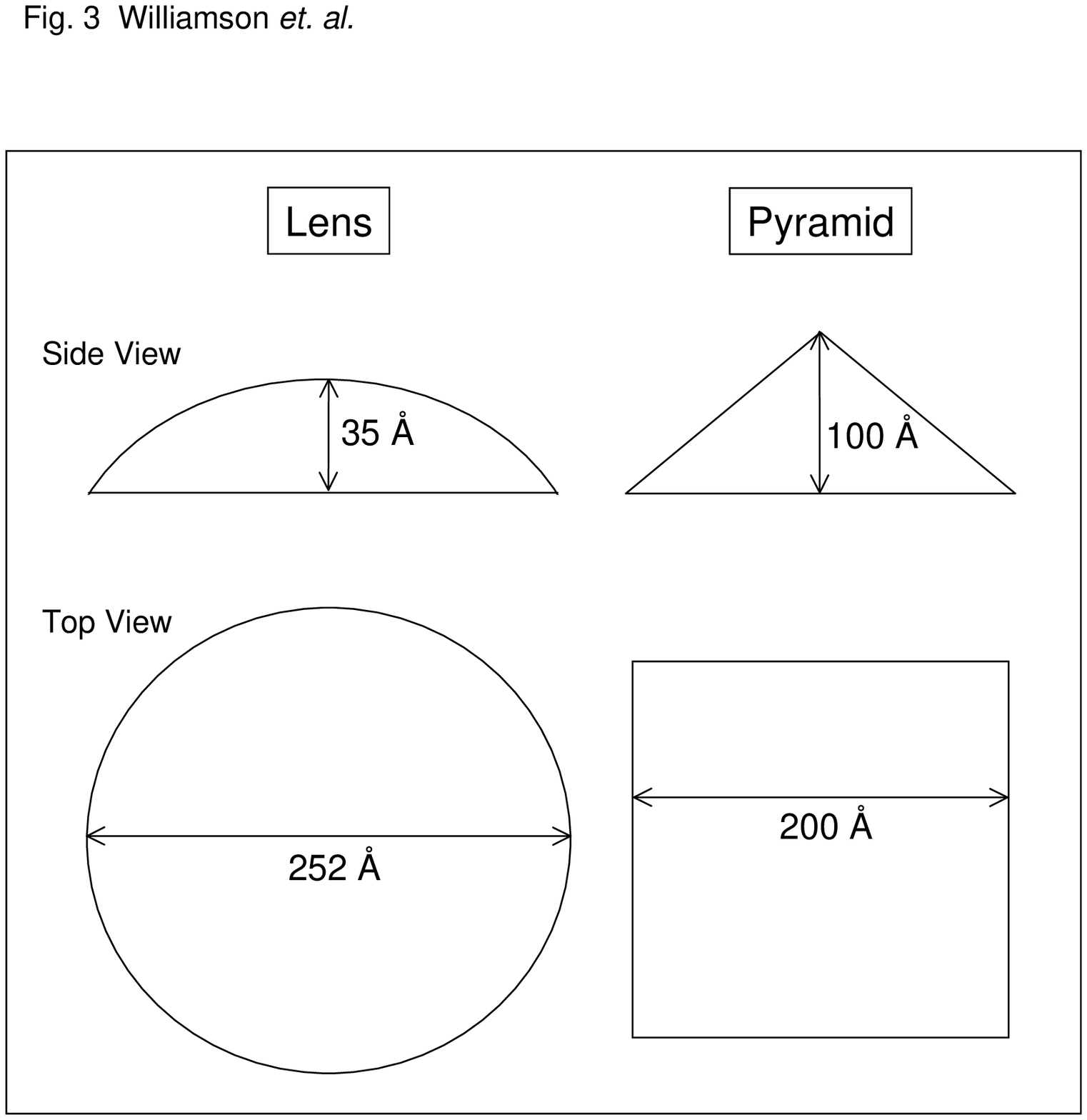}
\caption{Assumed model geometry of the lens and pyramidal shaped
quantum dots.}
\label{geometry}
\end{figure}

The electronic structure of a series of GaInAs/GaAs self-assembled
quantum dots was calculated using the methodology described in
Section~\ref{methods}.  We have chosen to focus on the well
established ``lens shaped'' dot geometry from
Refs.\cite{drexler94,medeiros95,fricke96,miller97,warburton97,schmidt97,warburton98,schmidt98}.
The shape of this dot is shown in Fig.~\ref{geometry}.  The profile is
obtained by selecting the section of a pure InAs sphere that yields a
circular base with diameter 252\AA~and a height of 35 \AA.  The main
experimental uncertainty about this dot is the composition profile.
It is not known if the dots are pure InAs or if Ga has diffused into
the dots.  For comparison, we also calculate the electronic structure
of a square based InAs pyramid with a base of 113\AA~and a height of
56\AA.  This is not believed to be a realistic geometry, however, it
has been used as a benchmark for many previous theoretical
calculations\cite{wang2000,jkim98,williamson99,grundman95,jaros96} and
we include it here for comparison purposes.  In the following sections
these two geometries will be referred to as the ``lens'' and the
``pyramid''.  The results of our calculations are shown in
Table~\ref{results_table} and Fig.~\ref{wavefuns}.

\subsection{Confined electron states}
Figure~\ref{wavefuns} shows the calculated square of the envelope
function for the electron states in the pyramidal and lens shaped
InAs/GaAs quantum dots.  For the lens shaped dot, the electron states
can be approximately interpreted as eigenstates of the $\hat{L_z}$
operator\cite{hawrylak}.  Here we plot only the first 6 bound states
corresponding to $l_z=0,\pm 1$ and $\pm 2$.  The first state $e_0$,
has $l_z=0$ and is commonly described as $s$-like as it has no nodes.
The $e_1$ and $e_2$ states have $l_z=\pm 1$, and are $p$-like with
nodal planes (110) and ($\overline{1}$10).  The $e_3,e_4$ and $e_5$
states have $l_z=\pm 2$ and 0 respectively and are commonly described
as $d_{x^2-y^2}$, $d_{xy}$ and $2s$ respectively.  Due to the
underlying zincblende atomistic structure, the $C_\infty$ symmetry is
reduced to $C_{2v}$.  Hence, the $e_0$ to $e_5$ states correspond to
the $a_1,b_1,b_2,a_1,a_2$ and $a_1$ irreducible representations of the
$C_{2v}$ group, rather than eigenstates of $\hat{L_z}$.  This allows
states $e_0,e_3$ and $e_5$ to couple.  This coupling is evident, for
example, in the larger charge density along [110] compared to
[1$\overline{1}$0] in the $e_3$ state, due to its coupling with $e_1$.
The observable effect of this $C_{2v}$ symmetry is to split the $e_1$
and $e_2$ $p$-states, $\delta_{pp}$, and the $e_3$ and $e_4$
$d$-states, $\delta_{dd}$.  The alignment of the $e_1$ and $e_2$
$p$-states states along the [110] and [1$\overline{1}$0] directions
also results from the underlying zincblende lattice structure.  Note,
this analysis neglects the effects of the spin-orbit interaction which
reduces the $C_{2v}$ group to a double group with the same single
representation for all the states.  In our calculations the spin-orbit
interaction is included, but is produces no significant effects for
the electron states.

The electron states in the pyramidal dot also belong to the $C_{2v}$
group and show a one-to-one correspondence with those in the lens
shaped dot.  However, there are only 5 bound states in the pyramidal
dot due to its smaller size.  Here we define an electron state as
bound if its energy is below that of the unstrained, bulk GaAs
conduction band edge.

The calculated values of the $s$-$p$ and $p$-$d$ energy spacings,
$\delta_{sp}$, and, $\delta_{pd}$, for the lens and pyramidal shaped
dots are 65 and 68 meV and 108 and 64 meV respectively.  The splitting
of the two $p$ states, $\delta_{pp}=e_2-e_1$ are 2 and 26 meV
respectively.  The calculated values of the electron binding energy,
$\Delta E(e)$, are 271 and 171 meV respectively.  The
electron-electron direct Coulomb energies, $J^{ee}_{e_0e_0}$,
$J^{ee}_{e_1e_1}$ and $J^{ee}_{e_0e_1}$ in the lens and pyramidal dots
are calculated as 32, 25 and 25 meV and 40, 35 and 36 respectively.
On applying a magnetic field in the growth direction, we calculate an
increase in the splitting of the two $p$ states ($e_2-e_1$) in the
lens shaped dot from 2 to 20 meV.  Details of this magnetic field
calculation will be given in a future publication\cite{shumway2000}.
Finally, the energy of the electron wetting layer level, $\Delta
E_{WL}^{(e)}$, with thicknesses of 1 and 2 ML is 15 and 24 meV below
the CBM of unstrained bulk GaAs.

\subsection{Confined hole states}
\begin{figure}[h]
\includegraphics[width=\linewidth]{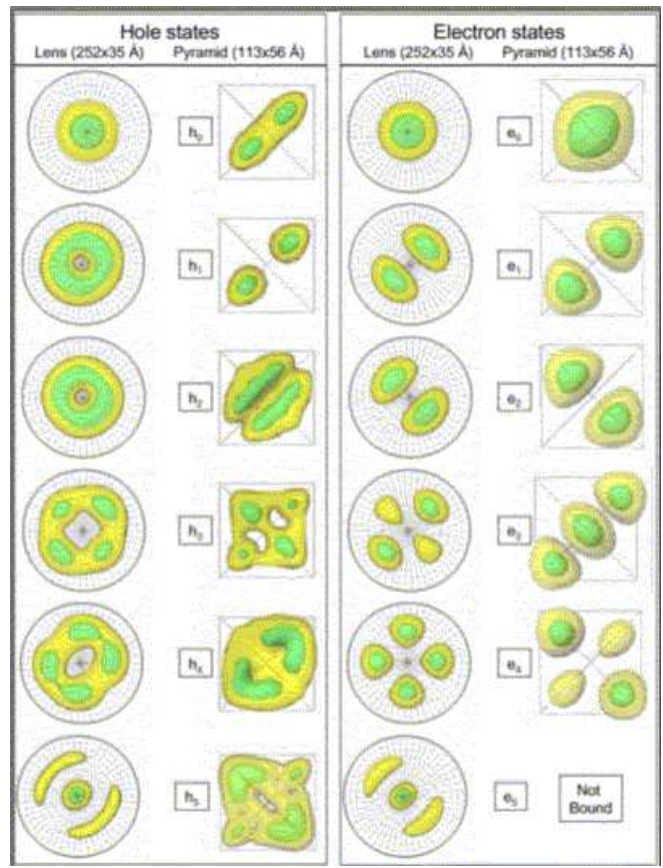}
\caption{Top view of the calculated electron and hole wavefunctions
squared for lens and pyramidal shaped InAs quantum dots embedded in
GaAs, with bases of 252 and 113 \AA~and heights of 25 and 56 \AA.  The
yellow and green isosurfaces represent 20 and 60 \% of the maximum
charge density.}
\label{wavefuns}
\end{figure}

Figure~\ref{wavefuns} shows calculated wavefunctions squared for the
hole states in pyramidal and lens shaped InAs/GaAs quantum dots.
Unlike the electron states, the hole states cannot be approximated by
the solutions of a single band Hamiltonian.  Instead there is a strong
mixing between the original bulk Bloch states with $\Gamma_{8v}$ and
$\Gamma_{7v}$ symmetry.  The larger effective mass for holes results
in a reduced quantum confinement of the hole states and consequently
many more bound hole states.  Only the 6 bound hole states with the
highest energy are shown in Figure~\ref{wavefuns}.  

The calculated values of the $h_0$-$h_1$ , $h_1$ -$h_2$ and
$h_2$-$h_3$ hole level spacings for the pyramidal and lens shaped dots
are 8,7 and 6 meV and 15, 20 and 1 meV respectively.  The calculated
hole binding energies, $\Delta E(e)$, are 194 and 198 meV.
We calculate the highest energy hole level in pure InAs wetting
layers, $\Delta E_{WL}^{(h)}$, with thicknesses of 1 and 2 ML to
reside 30 and 50 meV above the VBM of unstrained bulk GaAs.  The
hole-hole Coulomb energies, $J^{hh}_{h_0h_0}$, are 25 and 31 meV.

\subsection{Electron-hole excitonic recombination}
\begin{figure}[h]
\includegraphics[width=\linewidth]{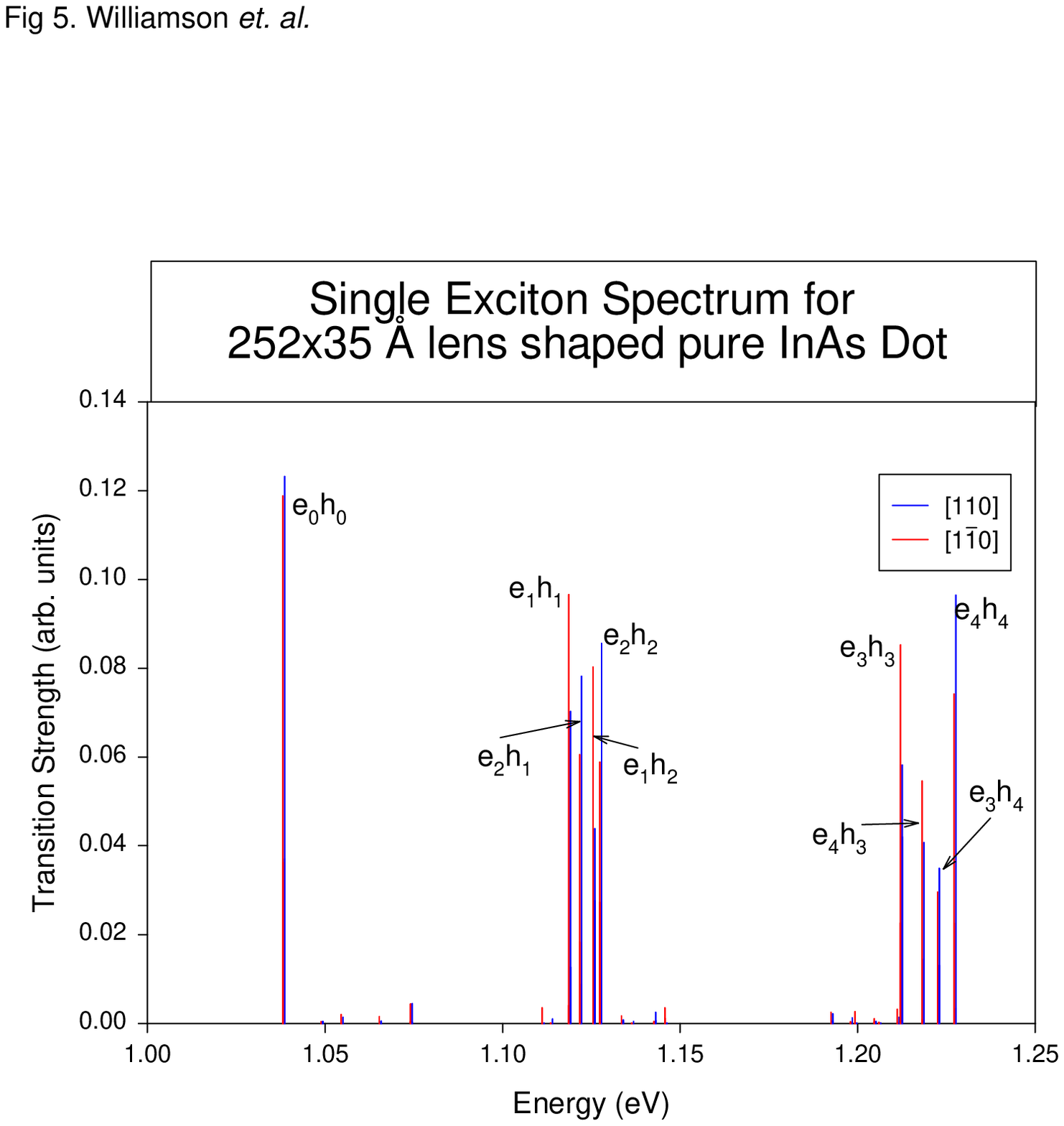}
\caption{The single exciton absorption spectrum for a pure InAs, lens
shaped dot with a base of 252 \AA~and a height of 35 \AA.  The
absorption peaks are calculated from Eq.(\ref{exciton}).  The ratios
of the dipole matrix elements for light polarized along [110] and
[1$\overline{1}$0] are calculated from Eq.(\ref{lambda}).}
\label{spectra}
\end{figure}

Figure~\ref{spectra} shows our calculated single exciton absorption
spectrum for a pure InAs, lens shaped dot with a base of 252 \AA~and a
height of 35 \AA.  The energies of each of the absorption peaks are
calculated from Eq.(\ref{exciton}).  The ratios of the dipole matrix
elements for light polarized along [110] and [1$\overline{1}$0] are
calculated from Eq.(\ref{lambda}).  Figure~\ref{spectra}, illustrates
that, for a lens shaped dot, both the conventional $e_i\rightarrow
h_i$ transitions and additional, $e_1-h_2, e_2-h_1, e_3-h_4$ and
$e_4-h_3$ transitions are strongly allowed.  The ratio of the
polarization anisotropies, $\lambda$, are shown in
Table~\ref{polarization_table}.  As a result of the circular symmetry
of the lens shaped dot, we calculate a polarization ratio of
$\lambda=1.03$ for the $e_0-h_0$ transition.  This value is in
contrast to that calculated value for a pyramidal dot of
$\lambda=1.2$\cite{wang99:2}.  For the higher angular momentum
transitions we find larger deviations from unity.  The magnitude of
the ratios, follows the polarization of the wavefunctions shown in
Fig.~\ref{wavefuns}.  For example we find ratios greater and then less
than unity for the $e_1-h_1$ and $e_2-h_2$ transitions, as reflected
by the elongations of the $e_1,h_1$ and $e_2,h_2$ wavefunctions along
the [110] and [1$\overline{1}$0] directions.
\begin{table}[hbt]
\caption{Calculated polarization anisotropy,
$\lambda=P_{110}:P_{1\overline{1}0}$, for lens shaped and pyramidal
Ga$_x$In$_{1-x}$As quantum dots embedded within GaAs.}
\label{polarization_table}
\begin{tabular*}{\linewidth}{@{\extracolsep{\fill}}ccc}
\toprule
& {Lens} & Pyramid \\
Geometry & 252x35\AA & 200x100\AA \\
\% Ga at base,tip,average & 0,0,0 & 0,0,0 \\
\colrule
$e_0-h_0$ & 1.03 & 1.20 \\
$e_1-h_1$ & 0.82 & 2.40 \\
$e_2-h_2$ & 1.27 & 0.52 \\
$e_3-h_3$ & 0.73 & 4.26 \\
$e_4-h_4$ & 1.23 & 0.63 \\
\botrule
\end{tabular*}
\end{table}

We calculate ground state electron-hole direct Coulomb energies,
$J^{eh}_{e_0h_0}$, of 37 and 31 meV in the lens shaped and pyramidal
dots.  The calculated ground state electron-hole exchange energies,
$K^{eh}_{e_0h_0}$ are an order of magnitude smaller, with values of 3
and 0.2 meV.  These yield excitonic band gaps of 1.03 and 1.12
respectively.  The calculated polarization anisotropy ratios
[Eq.(\ref{lambda})] for light polarized along [110] and
[1$\overline{1}$0] directions are $\lambda=1.03$ and 1.20 for the lens
and pyramidal shapes respectively.  The calculated excitonic dipoles
[Eq.(\ref{dipole})] are -3.1 and 0.16\AA~respectively.  A positive
dipole is defined as the center of the hole wavefunction being located
above the center of the electron wavefunction.

\section{Analysis of pertinent experimental measurements}\label{expt}

\subsection{The intra-band $s$-$p$ and $p$-$d$ electron energy spacings}\label{delta_sp}
Measurements of the spacing between the $e_0$ and $e_1$ like electron
levels ($s$-like and $p$-like) are based on infra red absorption.  For
the lens shaped dots, Fricke {\em et. al.}\cite{fricke96} load
electrons into the dots by growing a sample consisting of a n-type
doped layer, a tunneling barrier, a layer of InAs/GaAs lens shaped
dots, a GaAs spacer and a GaAs/AlAs short period superlattice (SPS).
By applying a voltage between the n-doped layer at the bottom of the
sample and a Cr contact grown on top of the SPS, electrons are
attracted from the n-doped layer into the InAs dots.  Infra-red
photons were used to excite electrons from the occupied $e_0$ level
into the $e_1$ level.  Neglecting the small exchange energy, the
energy differences for the $e_1-e_0$ excitations when 1 and 2
electrons are present in the dot are
\begin{eqnarray}\label{ir}
E_{01}[e_1^1]-E_{01}[e_0^1] & = & \left( \epsilon_{e_1} -
\epsilon_{e_0}\right) \nonumber \\
E_{02}[e_0^1e_1^1]-E_{02}[e_0^2] & = & \left( \epsilon_{e_1} -
\epsilon_{e_0} \right) + \left[ J^{ee}_{e_1,e_0} - J^{ee}_{e_0,e_0}
\right] \;\;\;,
\end{eqnarray}
The first of these energy differences yields a direct measurement of
the $s-p$ energy spacing, $\delta_{sp}$, of 49.1 meV.  The second
energy difference was measured at 50.1 meV.  Drexler {\em
et. al.}\cite{drexler94} also used infra red transmission spectroscopy
to measure the energy spacing, $\delta_{sp}=41$ meV.  Pan {\em
et. al.}\cite{pan98,pan98:2} have also performed infra-red absorption
measurements on truncated pyramidal dots with a base of 180 \AA~and
height of $\sim$60 \AA.  In these experiments, no gate voltage is
applied, and therefore the excitations take place from the ground
state of the samples, $E_{00}$.  They observe multiple infra-red
absorption peaks between 89 and 103 meV.  These could be associated
either with the $s$-$p$ spacing of the electron levels or spacings of
the hole states (see below).

Itskevich {\em et. al.}\cite{itskevich99} perform high power PL
measurements of pyramidal dots with a base of 150 \AA~and a height of
30 \AA~.  This high power excitation is able to simultaneously load
multiple excitons into the dots.  Due to state filling, these multiple
excitons will occupy ground state ($e_0,h_0$) and higher 
($e_1,e_2,h_0,h_1,h_2$) single particle levels.  Therefore the PL
measurements are able to simultaneously measure recombination between
electrons occupying the $e_0, e_1, e_2$ and $e_3$ levels with holes in
the $h_0, h_1, h_2$ and $h_3$ levels.  In general, to describe the
total energy differences associated with decay from $N$ to $N-1$
excitons occupying a dot requires a treatment that includes the
exchange and correlation between multiple occupational configurations
of the $N$ and $N-1$ excitonic states.  Such a configurational
interaction (CI) approach has previously been considered for model
parabolic dots\cite{hawrylak} and will be discussed for realistic dots in
a future publication\cite{williamson2000:2}.  For the purposes of this
discussion, we limit ourselves to discussing the energy differences
associated with the lowest energy configurations on the $N$ exciton
state, i.e. those predicted by the Aufbau principle.  Within this
approximation, the peaks in the high power, PL spectra can be
interpreted as corresponding to [see Eq.(\ref{ee1}) where exchange is
neglected]
\begin{eqnarray}\label{its}
\mbox{Peak 1:} \;\;\; && E_{11}[h_0^1e_0^1]-E_{00} = \left(
\epsilon_{e_0} - \epsilon_{h_0} \right) - J^{eh}_{e_0,h_0} \nonumber
\\ \mbox{Peak 2:} \;\;\; &&
E_{33}[h_0^2h_1^1e_0^2e_1^1]-E_{22}[h_0^2e_0^2] = \left(
\epsilon_{e_1} - \epsilon_{h_1} \right) - J^{eh}_{e_1,h_1} \nonumber
\\ && + 2\left[ -J^{eh}_{e_1,h_0} - J^{eh}_{e_0,h_1} + J^{ee}_{e_1,e_0}
+ J^{hh}_{h_1,h_0}\right] \nonumber \\ \mbox{Peak 3:} \;\;\; &&
E_{77}[h_0^2h_1^2h_2^2h_3^1e_0^2e_1^2e_2^2e_3^1]-E_{66}[h_0^2h_1^2h_2^2e_0^2e_1^2e_2^2]\;\;
\nonumber \\ &&  = \left( \epsilon_{e_3} - \epsilon_{h_3} \right) - J^{eh}_{e_3,h_3}
\nonumber \\ && + 2\sum_{i=0}^{2}\left[-J^{eh}_{e_3,h_i} -
J^{eh}_{e_i,h_3} + J^{ee}_{e_3,e_i} + J^{hh}_{h_3,h_i} \right] \;\;\;
.
\end{eqnarray}
Note, peak 3 is not assigned to a recombination from $e_2$ to $h_2$ as
this is almost degenerate with peak 2. Itskevich {\em et. al.}  assume
that (i) the Coulomb integrals in the square brackets on the right
hand side of Eq.(\ref{its}) cancel, (ii) that
$J^{eh}_{e_0,h_0}=J^{eh}_{e_1,h_1}=J^{eh}_{e_3,h_3}$, and (iii) that
the hole spacings, $\delta_{h_0h_1}, \delta_{h_1h_2}$, are small
compared to the electron level spacings.  With these assumptions the
spacings between peaks 1 and 2 and peaks 2 and 3 can be assigned to
the $s$-$p$ and $p$-$d$ energy spacings.  They find spacings
$\delta_{sp}$ and $\delta_{pd}$ of 75 and 80 meV respectively.  Our
calculations suggest that assumptions (i), (ii) and (iii) probably
introduce errors of $\sim\pm$10, $\sim\pm$5 and +10 meV respectively.
The neglect of exchange interactions in the above discussion also
introduces an error of  $\sim\pm$5 meV.

\subsection{The intra-band electron $p$ level splitting}
For the lens shaped dots, the Capacitance Voltage spectroscopy of
Fricke {\em et. al.}\cite{fricke96} can be used to estimate the
splitting of the $p$ states, $\delta_{pp}$, by loading two electrons
into the dot and exciting them using FIR spectroscopy.  The relevant
energy differences are
\begin{eqnarray}
E_{02}[e_0^1e_1^1]-E_{02}[e_0^2] & = & \left( \epsilon_{e_1} -
\epsilon_{e_0} \right) + \left[ J^{ee}_{e_1,e_0} - J^{ee}_{e_0,e_0}
\right] \nonumber \\
E_{02}[e_0^1e_2^1]-E_{02}[e_0^2] & = & \left( \epsilon_{e_2} -
\epsilon_{e_0} \right) + \left[ J^{ee}_{e_2,e_0} - J^{ee}_{e_0,e_0}
\right] \;\;\;.
\end{eqnarray}
By assuming $J^{ee}_{e_1,e_0}=J^{ee}_{e_2,e_0}$, the difference in the
two above expressions yields the energy spacing $\epsilon_{e_2} -
\epsilon_{e_1}$. They find a value of $\sim$2 meV. To measure the
effect of a magnetic field on the splitting of the $p$ states Fricke
{\em et. al.}\cite{fricke96} use infra-red absorption to measure the
above energy differences in an applied magnetic field.  At a field of
15 Tesla they measure an energy spacing of 19 meV.  More
theoretically, Dekel {\em et. al}\cite{dekel98} have demonstrated that
one must assume a splitting of the $p$ states to explain the number of
multi-exciton levels observed in their single dot, micro PL
measurements.

\subsection{The intra-band hole energy spacings}
There are currently no measurements available for the energy spacings
between the hole states in the lens shaped dots.  Itskevich {\em
et. al.}\cite{itskevich99} have performed high power PL measurements
of dots estimated to be square based, truncated pyramids with a base
of 150\AA~and a height of 30\AA~, under hydrostatic pressure to
estimate hole level spacings.  At hydrostatic pressures above 55 kbar,
they measure the PL associated with transitions from the
$X_{1c}$-state in the GaAs matrix to the $h_0$ and $h_1$ levels in the
dots.  They estimate a hole level spacing of $\sim$15 meV.  However,
the nature of the initial and final hole states is unclear.

Sauvage {\em et. al.}\cite{sauvage99} use polarized photoinduced
intraband absorption spectroscopy to measure the energy spacing
between the lowest hole state, $h_{000}$, and the hole state with a
single node in the growth direction, $h_{001}$.  This corresponds to
[see Eq.(\ref{ee1})]
\begin{eqnarray}
E_{11}[h_{000}^1e_0^1]-E_{11}[h_{001}^1e_0^1] & = &\left(
\epsilon_{h_{001}} - \epsilon_{h_{000}} \right) \nonumber \\ & + &
\left[ J^{eh}_{e_0,h_{000}} - J^{eh}_{e_0,h_{001}}\right] \;\;\;.
\end{eqnarray}
By assuming that $J^{eh}_{e_0,h_{000}}=J^{eh}_{e_0,h_{001}}$ Sauvage
{\em et. al.}  estimate the $h_{001}-h_{000}$ spacing to be $\sim$120
meV.  Note $h_{001}$ is almost certainly higher in energy than states
with nodes perpendicular to the growth direction ($h_{010}$ and
$h_{100}$) due to the smaller dimension of the dot in the growth
direction.  Consequently, the energy difference $h_{001}-h_{000}$ is
not the spacing of the first two hole states $h_0$-$h_1$.

Tang {\em et. al.}\cite{tang98} measure activation energies for
excitations from $h_0$ and $h_1$ to the hole wetting layer of 48 and
30 meV respectively, implying an $h_0$-$h_1$ spacing of $\sim$18 meV.

\subsection{The electron and hole binding energies, $\Delta E(e)$ and
$\Delta E(h)$} There have been no direct measurements of the electron
or hole binding energy for lens shaped InAs dots.  However, it has
been measured in other dots by several groups using a range of
techniques.  Berryman {\em et. al.}\cite{berryman97} placed pyramidal
InAs dots estimated to have a base of 100 \AA~and height of 15 \AA~in
a p-n junction and measured the temperature dependence of the ac
conductance as a function of frequency.  These measurements predict a
hole binding, $\Delta E(h)$, energy of $\sim$240 meV.  When subtracted
from the bulk GaAs band gap, this yields a value for the electron
binding energy, $\Delta E(e)$, of $\sim$60 meV.  The authors obtain
similar results from temperature dependent Hall measurements of
thermal hole trapping.  Itskevich {\em et. al.}\cite{itskevich98}
measured the pressure at which PL measurements could detect a
$\Gamma$-X crossing in pyramidal InAs/GaAs quantum dot samples.  By
extracting these PL measurements back to zero pressure they were able
to extrapolate a value for the electron binding energy, $\Delta E(e)$,
of $\sim$50 meV.  Itskevich {\em et. al.}\cite{itskevich99} also used
high pressure PL to measure the energy difference between the $X_{1c}$
level in bulk GaAs and the $h_0$ level in the quantum dots.  By
extrapolating this value back to zero pressure, they predict a value
for the hole binding energy, $\Delta E(h)$, of $\sim$250 meV.  Brunkov
{\em et. al.}\cite{brunkov98} performed capacitance-voltage
spectroscopy measurements, which when fitted to a capacitance model
for their dot geometry predict an electron binding energy of 80 meV
for dots with bases of 250 \AA.  Tang {\em et. al.}\cite{tang98}
measured the spacing of the electron wetting layer to both the GaAs
CBM and the $e_0$ level and hence deduced a value for the electron
binding energy, $\Delta E(e)$, of $\sim$80 meV, for dots with an
estimated base of 130 to 170 \AA.

\subsection{The position of the electron and hole wetting layer level}
The presence of a distinct wetting layer signal in the PL spectra of a
sample of self assembled quantum dots is the hallmark of a high
quality sample.  In samples where the wetting layer has ``dissolved''
due to the growth conditions, it is likely that the geometry and
composition of the quantum dots will also have dramatically altered
from their uncapped state.

In the lens shaped InAs dots Schmidt {\em et. al.}\cite{schmidt97}
observe PL emission from the ground state of the wetting layer at 1.34
eV.  Photovoltage measurements\cite{schmidt97} on the same samples
show a strong peak corresponding to absorption into the ground state
of the wetting layer at 1.35 eV.  There are currently no measurements
available for the position of the individual electron and hole wetting
layers in the lens shaped InAs/GaAs quantum dot samples.  In
Ref.\cite{sauvage97} Sauvage {\em et. al.} grow lens shaped InAs
dots with an estimated base of 150 \AA~and a height of 30 \AA~on a
substrate that is n-doped with silicon.  This n-doping loads electrons
into the $e_0$ state in the dot, which they excite into the wetting
layer using infra-red excitation.  In these samples they estimate the
wetting layer to be 150 meV above the $e_0$ level.  In
Ref.~\cite{sauvage99} Sauvage {\em et. al.} load electrons into
the $e_0$ state of similar dots using an optical interband pump.
Subsequent infra-red absorption places the wetting layer 190 meV above
the $e_0$ state.  Tang {\em et. al.}\cite{tang98} measure thermal
transfer of holes to the wetting layer, and obtain a spacing between
the $h_0$ level and the hole wetting layer, $\Delta E_{WL}^{(e)}$ of
$\sim$48 meV.  They also measure thermal transfer from an excited
state, possibly involving $h_1$, which places the hole wetting layer
$\sim$30 meV below the $h_1$ level.

\subsection{The number of confined electron and hole states}
The actual number of confined electron and hole states in a self
assembled InAs/GaAs quantum dot depends on the size and composition of
the dot.  Early single band, effective mass
calculations\cite{grundman95} for pure InAs pyramidal dots with a base
of 120 \AA~and height 60 \AA~predicted only a single bound electron
state and several bound hole states.  Consequently, many experiments
were then interpreted in this light.  More accurate multi-band
$k.p$\cite{stier99,jiang97,jaros96} and
pseudopotential\cite{jkim98,williamson99} calculations have predicted
5 or more bound electron states in the same dots.

The high power PL experiments of Itskevich {\em
et. al.}\cite{itskevich99} show the gradual disappearance of 5 peaks
as a function of external pressure.  This is interpreted as direct
evidence for 5 confined electron levels in their samples.  The single
dot, multi-exciton measurements of Dekel {\em et. al.}\cite{dekel98}
require the assumption of at least 3 bound electron states to explain
their experimental spectra.  Similarly, the capacitance-voltage
spectroscopy of Fricke {\em et. al.}\cite{fricke96} shows two peaks
corresponding to the capacitance of $s$-like states and the nearly
degenerate $p$-like states, providing evidence for at least 3 bound
electron states.

\subsection{Electron and hole Coulomb and Exchange Interactions}\label{ee_repulsion}
By loading multiple electrons and holes into quantum dots it is
possible to measure the Coulomb and exchange interactions between
these additional electrons and holes.  The magnitude of these
interactions is a function of the shape of the electronic
wavefunctions (see section~\ref{methods}) and provides an additional
quantity to test the accuracy of theoretical models.

To study electron-electron interactions, Fricke {\em
et. al.}\cite{fricke96} use the same experimental setup discussed in
section~\ref{delta_sp}.  From Eq.(\ref{ee1}) we see the energy
differences corresponding to the peaks in the Capacitance Voltage (CV)
spectra associated with loading one and two electrons into the $e_0$
level in the dots is
\begin{eqnarray}
E_{01}\;[e_0^1] - E_{00}\;      & = & \epsilon_{e_0} \nonumber \\
E_{02}\;[e_0^2] - E_{01}\;[e_0^1] & = & \epsilon_{e_0} + J^{ee}_{e_0,e_0} \;\;.
\end{eqnarray}
The electron-electron Coulomb interaction, $J^{ee}_{e_0,e_0}$, can
therefore be directly measured as the splitting of these two CV peaks.
They find a value of $J^{ee}_{e_0,e_0}\sim$23 meV.  From Eq.(\ref{ir})
we see that
\begin{equation}
J^{ee}_{e_0,e_1}=J^{ee}_{e_0,e_0}+(50.1-49.1) = 24 \mbox{meV} \;\;\;.
\end{equation}
Finally by fitting 4 equidistant bell curves to the CV spectra
corresponding to loading 3,4,5 and 6 electrons into the dots, an
approximate value for the charging energy between the $p$ states,
$J^{ee}_{e_1,e_1}$, of $\sim$18 meV is obtained.  From Eq.(\ref{ee1})
we that the spacing
$E_{04}-E_{03}=J^{ee}_{e_1,e_1}+2J^{ee}_{e_0,e_1}$, while
$E_{06}-E_{05}=3J^{ee}_{e_1,e_1}+2J^{ee}_{e_0,e_1}$ and hence the
approximation of equidistant peaks will introduce some error into this
estimate for $J^{ee}_{e_1,e_1}$.

\subsection{Electron-hole excitonic recombination}
To our knowledge there have so far been no measurements of the
polarization anisotropy in the lens shaped dots discussed here.  The
polarization anisotropy for $e_i-h_i$ excitonic recombination in
InAs/GaAs was measured by Yang {\em et. al.}\cite{yang99,yang2000} ,
who find a ratio of $\lambda_{e_0,h_0}=1.2$ and
$\lambda_{e_2,h_2}=1.3$ for InAs dots whose geometry is measured to be
formed by four \{136\} faceted planes with bases ranging from 150 to
250 \AA~and a base to height ratio of 4:1. Yang {\em
et. al.}\cite{yang99,yang2000} have performed k.p calculations for
this dot geometry, which include the ``geometric factor'' but not the
``atomic symmetry factor'' discussed in section~\ref{quantities}. They
find $\lambda_{e_0,h_0}=1.8$ and $\lambda_{e_2,h_2}=3.5$.  The authors
suggest the k.p simulations of the measured polarization ratio can be
used to deduce the geometric shape anisotropy.  However, we
demonstrate here that when the ``atomic symmetry factor'' is included
an anisotropy of $\lambda_{e_0,h_0}=1.3$ is obtained even for a {\em
square} based pyramid.  Thus k.p simulations lacking the ``atomic
symmetry'' factor can not be used to reliably deduce the geometric
shape anisotropy.

To our knowledge there have so far been no measurements of the
excitonic dipole in lens shaped InAs dots.  Fry {\em
et. al.}\cite{fry2000} used photocurrent spectroscopy within an
applied electric field to measure the excitonic dipole moment,
$d_{h_i,e_j}$, [Eq.(\ref{dipole})].  They find the center of the hole
wavefunctions to be located $\sim$4 \AA~above the center of the
electron wavefunction (positive dipole). Fry {\em
et. al.}\cite{fry2000} also perform single-band, effective mass
calculations, in an attempt to isolate the origin of this dipole.
They predict that in the absence of alloying the dipole is -3 \AA,
i.e. the opposite sign , but a linear composition profile with
Ga$_{0.5}$In$_{0.5}$As at the base and pure InAs at the top of a
truncated pyramid with a base of 155 \AA, and height 55 \AA,
reproduces the correct dipole of 4 \AA.  They suggest that this
alloying profile explains the observed dipole.  We have repeated these
calculations and confirm that, within a single-band model, such a
composition profile causes both electrons and holes to move up in the
dot compared to their positions in a pure InAs dot.  The heavier
effective mass of the holes, results in less kinetic energy associated
with confinement at the top of the dot and hence the holes move up
more than electrons on the introduction of Ga, producing the correct
dipole.  However, when we repeat these calculations in the more
sophisticated, multi-band LCBB basis used here, we find significant
heavy hole-light hole mixing in the $h_0$ state, which acts to reduce
the above effect and produce a smaller dipole of $\sim$1 \AA, in
contradiction with experiment (+4 \AA).  We therefore conclude that in
a more complete calculation the shape and alloy profile postulated by
Fry {\em et. al.}\cite{fry2000} does not produce the observed
excitonic dipole.

\section{Comparison of experiment and theory}

In Table~\ref{results_table} we show the results of our calculations
for pure InAs, lens shaped quantum dots embedded within GaAs [column
(a)].  Table~\ref{results_table} also shows the experimentally
measured splittings of the electron levels, the electron-electron and
electron-hole Coulomb energies, the magnetic field dependence and the
excitonic band gap measured in Refs.\cite{fricke96} and
\cite{warburton98}.  The agreement between the measured energy
level spacings, Coulomb energies and magnetic field response with our
theoretical lens shaped model is generally good.  Both the model and
experiment find (i) a large spacing, $\delta_{sp}$, ($\sim$50-60 meV)
between the $s$-like $e_0$ state and the $p$-like $e_1$ state, (ii) a
small spacing, $\delta_{pp}$, ($\sim$3 meV) between the two $p$-like
$e_1$ and $e_2$ states and (iii) a large spacing ($\sim$55 meV)
between the $p$-like $e_2$ state and the $d$-like $e_3$ state.

These electron level spacings are similar to those found for pyramidal
quantum dots\cite{jkim98} (see Table~\ref{results_table}).  However,
in the pyramidal dot, the spacings of the two $p$-like and $d$-like
states, $\delta_{pp},\delta_{dd}$, is larger (26 and 23 meV) as a
result of the lower C$_{2v}$ symmetry of a zincblende pyramid.  Both
the model and experiment also find similar values for the Coulomb
energies, $J(e_0e_0)$ and $J(e_0h_0)$ ($\sim$25 meV).  

The calculated hole binding energy of $\Delta E(h)=193$ meV is in good
agreement with those of Berryman {\em et. al.}\cite{berryman97}
($\sim$240 meV) and Itskevich {\em et. al.}\cite{itskevich99}
($\sim$250 meV).  Our calculated electron binding energies, $\Delta
E(h)$, are considerable larger (271 meV) than those of Berryman {\em
et. al.}\cite{berryman97} ($\sim$60 meV) and Tang {\em
et. al.}\cite{tang98} ($\sim$80 meV).  We attribute this difference to
the larger size of our dots. The assumption of a pure InAs dot also
affects the comparison.  The agreement improves when we include Ga in
our dots (see section~\ref{diffusion}).

The calculated electron-electron and electron-hole Coulomb energies
are in reasonable agreement with those extracted from
Refs.\cite{fricke96} and \cite{warburton98}.  For the
integrals $J_{e_0e_0}^{ee}, J_{e_0e_1}^{ee}, J_{e_1e_1}^{ee}$ and
$J_{e_0h_0}^{eh}$ we calculate values of 31, 25, 25 and 37
respectively, compared to measured values of 23, 24, 18 and 33.3 meV

The calculated polarization anisotropies, $\lambda$, for the $e_0-h_0$
recombination in lens and pyramidal shaped, pure InAs dots are
$\lambda=1.03$ and 1.2 respectively.  A future measurement of this
anisotropy ratio for lens shaped dots would provide an important
piece of evidence for determining the detailed geometry of the dots.

In the lens shaped dot we find a difference in the average positions
of the $h_0$ and $e_0$ states, $d_{h_i,e_j}$, of around 1 \AA.  This
is smaller than the value we calculate for a pyramidal quantum dot,
where we find the hole approximately 3.1 \AA~higher than the
electron.

In summary, the assumed lens shaped geometry, with a pure InAs
composition produces a good agreement with measured level splitting,
Coulomb energies and magnetic field dependence.  A closer inspection
of the remaining differences reveals that the calculations
systematically {\em overestimate} the splittings between the single
particle electron levels ($\delta_{sp}$: 65 {\em vs.}  50 meV,
$\delta_{pd}$:68 {\em vs.} 48 meV) and {\em underestimates} the excitonic
band gap (1032 {\em vs.}  1098 meV).

\subsection{Pure InAs dots: The effects of shape and size}
Focusing on the lens shape only, we examine the effect of changing the
height and base of the assumed geometry.  Calculations were performed
on similar lens shaped, pure InAs dots where (i) the base of the dot
was increased from 252 to 275\AA, while keeping the height fixed at
35\AA, [column (b)] and (ii) the height of the dot was decreased from
35 to 25\AA, while keeping the base fixed at 252\AA, [column (c)].  It
shows that decreasing the height of the dot increases the quantum
confinement and hence increases the splittings of the electron and
hole levels ($\delta_{sp}$: from 65 to 69 meV and $\delta_{h_0,h_1}$:
from 8 to 16 meV).  Decreasing the height of the dot also acts to
increase the excitonic band gap from 1032 to 1131 meV by pushing up
the energy of the electron levels and pushing down the hole levels.
Conversely, increasing the base of the dot decreases both the
splittings of the single particle levels ($\delta_{sp}$: from 66 to 61
meV) and the band gap (1032 to 1016 eV).  These small changes in the
geometry of the lens shaped dot have only a small effect on electronic
properties that depend on the shape of the wavefunctions.  The
electron-electron and electron-hole Coulomb energies remain relatively
unchanged, the magnetic field induced splitting remain at 20 meV, the
polarization anisotropy, $\lambda$, remains close to 1.0 and the
excitonic dipole, $d_{h_i,e_j}$, remains negligible.  In summary,
reducing either the height or the base of the dot increases quantum
confinement effects and hence increases energy spacings and band gaps,
while not significantly effecting the shape wavefunctions.

\subsection{Interdiffused In(Ga)As/GaAs lens shaped dots}\label{diffusion}
We next investigate the effect of changing the composition of the
quantum dots, while keeping the geometry fixed.  There have recently
been several experiments\cite{fry2000,metzger99,garcia97} suggesting
that a significant amount of Ga diffuses into the nominally pure InAs
quantum dots during the growth process.  We investigate two possible
mechanisms for this Ga in-diffusion; (i) Ga diffuses into the dots
during the growth process from all directions producing a dot with a
uniform Ga composition Ga$_x$In$_{1-x}$As, and (ii) Ga diffuses up
from the substrate, as suggested in Ref.\cite{fry2000}.  To
investigate the effects of these two methods of Ga in-diffusion on the
electronic structure of the dots, we compare pure InAs dots embedded
in GaAs with Ga$_x$In$_{1-x}$As, random alloy dots embedded in GaAs,
where the Ga composition, $x$, (i) is fixed at 0.15, [column (d)] and
(ii) varies linearly from 0.3 at the base to 0 at the top of the dot,
[column (e)].  

The electronic structure of these dots is compared in
Table~\ref{results_table}.  It shows that increasing the amount of Ga
in the dots acts to decrease the electron level spacings
($\delta_{sp}$: from 65 to 58 and 64 meV for $x=0.15$ and $x=0.3$ to
$x=0$ respectively).  It also acts to increase the excitonic band gap
from 1032 to 1080 and 1125 meV respectively.  The electron binding
energy, $\Delta E(e)$, is decreased by the in diffusion of Ga (from
271 to 209 and 192 meV), while the hole binding energy, $\Delta E(h)$,
is relatively unaffected.  This significant decrease in the electron
binding energy considerably improves the agreement with experiments on
other dot geometries\cite{berryman97,tang98}.  

As with changing the size of the dots, we find that Ga in-diffusion has
only a small effects on properties that depend on the shape of the
wavefunctions.  The calculated electron-electron and electron-hole
Coulomb energies are almost unchanged, while the average separation of
the electron and hole, $d_{h_i,e_j}$, increases from 0.16 to to 0.5
and 1.2 \AA~and the polarization ratio, $\lambda$, and magnetic field
response are also unchanged.

Table~\ref{results_table} shows that the dominant contribution to the
increase in the excitonic band gap and reduction in electron binding
energy, results mostly from an increase in the energy of the {\em
electron} levels as the Ga composition is increased.  This can be
understood by considering the electronic properties of the bulk
Ga$_x$In$_{1-x}$As random alloy.  The unstrained valence band offset
between GaAs and InAs is $\sim$ 50 meV\cite{wei98}, while the
conduction band offset in $\sim$ 1100 meV and hence changing the Ga
composition, $x$, has a large effect on the energy of the electron
states and only a small effect on the hole states.  In summary, the
effect of Ga in-diffusion is to reduce the spacing of the electron
levels while significantly increasing their energy and hence
increasing the band gap.  We find that only the average Ga composition
in the dots is important to their electronic properties.  Whether this
Ga is uniformly or linearly distributed throughout the dots has a
negligible effect.  Note, in Ref.\cite{fry2000} it is suggested
that a linear composition profile is required to produce an excitonic
dipole moment in agreement with that measured by stark experiments.
For the lens shaped geometry discussed here there have so far been no
such measurements of the dipole, but our calculations suggest that it
should be small ($\sim$1\AA).

\section{Discussion}
The effects of changing the {\em geometry} of the lens shaped, pure
InAs dots on the single particle energy levels can be qualitatively
understood from single band, effective mass arguments.  These predict
that decreasing any dimension of the dot, increases the quantum
confinement and hence the energy level spacings and the single
particle band gap will increase.  Note that as the dominant quantum
confinement in these systems arises from the vertical confinement of
the electron and hole wavefunctions, changing the height has a
stronger effect of the energy levels than changing the base.  In this
case decreasing the height by 10\AA~has a much stronger effect on the
energy spacings and on the band gap than increasing the base by 23\AA.

As increasing(decreasing) the dimensions of the dot acts to
decrease(increase) both the level spacings and the gap, it is clear
that changing the dot geometry alone will not significantly improve
the agreement with experiment as this requires a simultaneous {\em
decrease} in the energy level splittings and {\em increase} in the
band gap.  However, Ga in-diffusion into the dots acts to {\em
increase} the band gap of the dot while decreasing the energy level
spacings.  Table~\ref{results_table} shows that adopting a geometry
with a base of 275 \AA~ and a height of 35 \AA and a uniform Ga
composition of Ga$_{0.15}$In$_{0.85}$As produces the best fit to the
measurements in Refs.\cite{fricke96} and
\cite{warburton98}.  

In conclusion, our results strongly suggest that to obtain very
accurate agreement between theoretical models and experimental
measurements for lens shaped quantum dots, one needs to adopt a model
of the quantum dot that includes some Ga in-diffusion within the
quantum dot.  When Ga in-diffusion is included, we obtain an excellent
agreement between state of the art multi-band pseudopotential
calculations and experiments for a wide range of electronic
properties.  We are able to predict most observable properties to an
accuracy of $\pm 5$ meV, which is sufficient to make predictions of
both the geometry and composition of the dot samples.

\noindent{\bf Acknowledgments} We thank J. Shumway and
A. Franceschetti for many useful discussions and their comments on the
manuscript.  This work was supported DOE -- Basic Energy Sciences,
Division of Materials Science under contract No. DE-AC36-99GO10337.

\clearpage
\widetext
\begin{table}[hbt]
\caption{Calculated single particle electron and hole energy level
spacings, electron and hole binding energies, $\Delta E(e,h)$,
electron-electron and electron-hole Coulomb energies, 
excitonic band gap all in meV, exciton dipole moment and polarization anisotropy
for lens shaped and pyramidal Ga$_x$In$_{1-x}$As quantum dots embedded
within GaAs.}
\label{results_table}
\begin{tabular*}{\linewidth}{@{\extracolsep{\fill}}ccccccccc}
\toprule
&\multicolumn{6}{c}{Lens Calculations} & Pyramid Calc. & Lens Expt.\\
& (a) & (b) & (c) & (d) & (e) & (f) & (g) \\
Geometry & 252x35\AA & 275x35\AA & 252x25\AA & 252x35\AA & 252x35\AA &
275x35\AA & 200x100\AA & \cite{fricke96,warburton98} \\
\% Ga at base,tip,average & 0,0,0 & 0,0,0 & 0,0,0 & 15,15,15 & 30,0,15 & 15,15,15 & 0,0,0 \\
\colrule
$\delta_{sp}=e_1-e_0$ & 65  & 57  & 69  & 58  & 64  & 52  & 108 & 50   \\
$\delta_{pd}=e_3-e_2$ & 68  & 61  & 67  & 60  & 63  & 57  & 64  & 48   \\
$\delta_{pp}=e_2-e_1$ & 2   & 2   & 2   & 2   & 3   & 2   & 26  & 2    \\
$e_2-e_1$(15T)        & 20  & 20  & 18  & 21  & 20  & 17  &     & 19   \\
$\delta_{dd}=e_4-e_3$ & 4   & 3   & 4   & 4   & 3   & 1   & 23  &      \\
$h_0-h_1$             & 8   & 12  & 16  & 13  & 14  & 11  & 15  &      \\
$h_1-h_2$             & 7   & 6   & 5   & 5   & 6   & 5   & 20  &      \\
$h_2-h_3$             & 6   & 10  & 14  & 13  & 14  & 9   & 1   &      \\
$\Delta E(e)$         & 271 & 258 & 251 & 209 & 192 & 204 & 171 &      \\ 
$\Delta E(h)$         & 193 & 186 & 174 & 199 & 203 & 201 & 198 &      \\ 
$J_{e_0e_0}$          & 31  & 29  & 32  & 29  & 31  & 28  & 40  &  23   \\
$J_{e_0e_1}$          & 25  & 24  & 26  & 24  & 24  & 24  & 35  &  24   \\
$J_{e_1e_1}$          & 25  & 24  & 26  & 25  & 24  & 26  & 36  &  $\sim$18 \\
$J_{h_0h_0}$          & 30  & 27  & 39  & 32  & 28  & 30  & 31 \\
$J_{e_0h_0}$          & 30  & 28  & 35  & 31  & 29  & 29  & 31  & 33.3 \\
$K_{h_0e_0}$          & 0.15& 0.13& 0.14& 0.15& 0.1 & 0.12& 0.2 \\
$e_0-h_0-J_{e_0h_0}$ & 1032 & 1016 & 1131 & 1080 & 1125 & 1083 &
1127 & 1098 \\
$d_{e_0,h_0}$ (\AA)     & 0.16 & -0.37 & 0.5 & 0.5 & 1.2 & 0.5 & 3.1 \\
$\lambda=P_{110}:P_{1\overline{1}0}$ & 1.03 & 1.01  & 1.04 & 1.05 &
1.08 & 1.08 & 1.20 \\
\botrule
\end{tabular*}
\end{table}

\end{document}